\documentclass[a4paper,nofootinbib,twocolumn,pre,floatfix,showpacs,aps,10pt]{revtex4-1}

\usepackage{amsfonts,amsmath,amssymb}			
\usepackage{blindtext}
\usepackage{bm}
\usepackage{color}
\usepackage{dcolumn}
\usepackage{fancyhdr}	
\usepackage{float}
\usepackage{graphicx}
\usepackage[urlcolor=blue,colorlinks=true,pdfborderstyle={/S/U/W 1}]{hyperref}
\usepackage[latin1]{inputenc}	
\usepackage{natbib}
\usepackage{rotating}
\usepackage{siunitx}											
\usepackage{textcomp}
\usepackage{textgreek}
\usepackage{upgreek}
\usepackage{enumitem}
\usepackage[rightcaption]{sidecap}
\sidecaptionvpos{figure}{t}

\newcommand{\eg}{{e.g.}}
\newcommand{\etal}{\textit{et~al.}}

\newcommand{\Cm}{C_\mathrm{m}}
\newcommand{\cwa}{c_\mathrm{wa}}
\newcommand{\imconc}{N_\mathrm{S}}

\newcommand{\errz}{\sigma_z}

\newcommand{\GDPTprogram}{\emph{DefocusTracker}}

\newcommand{\Ncal}{N_\mathrm{cal}}
\newcommand{\Nsub}{N_\mathrm{sub}}

\newcommand{\detectedp}{\phi}

\newcommand{\beq}[1]{\begin{equation} \eqlab{#1}}
\newcommand{\eeq}{\end{equation}}
\newcommand{\bsub}{\begin{subequations}}
\newcommand{\esub}{\end{subequations}}
\def\bal#1\eal{\begin{align}#1\end{align}}
\def\bsubal#1\esubal{\bsub \begin{align}#1\end{align} \esub}

\newcommand{\eqlab}[1]{\label{eq:#1}}

\newcommand{\figref}[1]{Fig.~\ref{fig:#1}}

\newcommand{\figlab}[1]{\label{fig:#1}}

\newcommand{\secref}[1]{Section~\ref{sec:#1}}

\newcommand{\seclab}[1]{\label{sec:#1}}

\begin{document}

\title{A fast and robust algorithm for General Defocusing Particle Tracking}

\author{Massimiliano Rossi$^1$}
\email{rossi@fysik.dtu.dk}
\author{Rune Barnkob$^2$}
\email{rune.barnkob@tum.de}

\affiliation{
$^1$Department of Physics, Technical University of Denmark, DTU Physics Building 309, DK-2800 Kongens Lyngby, Denmark\\
$^2$Heinz-Nixdorf-Chair of Biomedical Electronics, Department
of Electrical and Computer Engineering, Technical
University of Munich, TranslaTUM, 81675 Munich, Germany}

\begin{abstract}
The increasing use of microfluidics in industrial, biomedical, and clinical applications requires a more and more precise control of the microfluidic flows and suspended particles or cells. This leads to higher demands in three-dimensional and automated particle tracking methods, \eg\ for use in feedback-control systems. General Defocusing Particle Tracking (GDPT) is a 3D particle tracking method based on defocused particle images which is easy to use and requires standard laboratory equipment. In this work, we describe in details a fast and robust algorithm for performing  GDPT, which is suitable for automatized and real-time applications. Its key feature is a fast, segmentation-free approach to identify particles and estimate their 3D position. This detection step is followed by a refinement and iteration step to improve accuracy and identification of overlapping particles. We show that the algorithm is versatile and can be applied to different types of images (darkfield and brightfield). We use synthetic image sets of varying particle concentration to evaluate the performance of the algorithm in terms of detected depth coordinate uncertainty, particle detection rate, and processing time. The algorithm is applied and validated on experimental images showing that it is robust towards background or illumination fluctuations. Finally, to test the algorithm on real-time applications, we use synthetic images to set up a simulation framework with experimentally-relevant parameters and where the true particle positions are known.
\end{abstract}

\maketitle

\begin{figure*}
    \centering
    \includegraphics[width=1.0\textwidth]{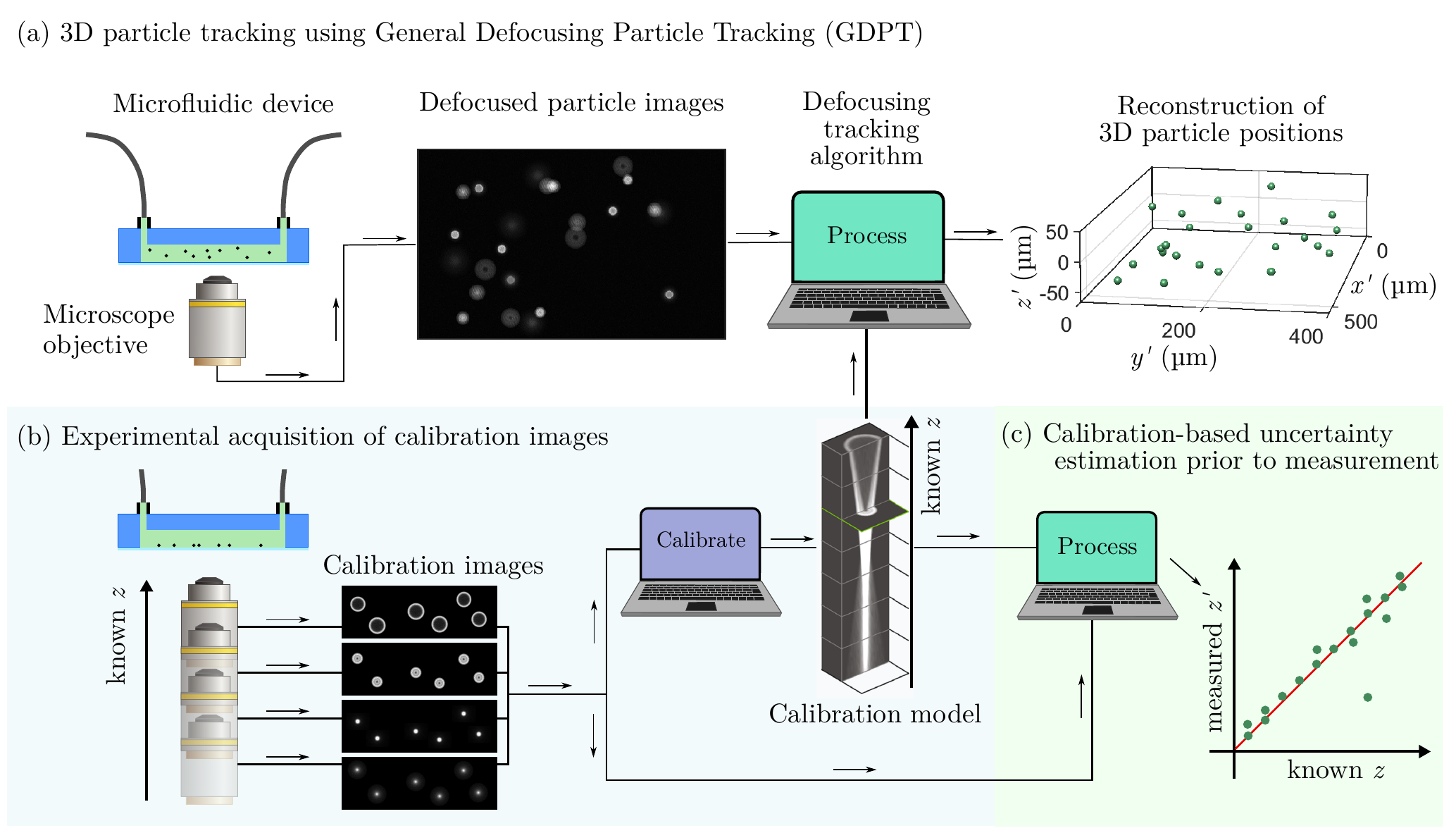}
	\caption{Example of single-camera, 3D particle tracking performed by General Defocusing Particle Tracking (GDPT). (a) Suspended microparticles are imaged with a conventional microscope resulting in 2D images with defocused particle images. The depth positions of the particles are determined by comparing their images to a reference set of already-acquired experimental particle images at known depth positions (calibration model). (b) The calibration model is obtained from a set of calibration images acquired experimentally by taking subsequent images of one or more reference particles displaced at known positions, e.g. by observing sedimented particles using a motorized microscope focusing stage. (c) The depth coordinate uncertainty and processing time can be estimated prior to the experiment by applying the tracking algorithm to the set of calibration images with the calibration model as input.}
	\figlab{intro}
\end{figure*}

\section{Introduction}

The recent advancements in microfluidic devices, specially in fields like biology or medicine, require more and more precise and continuous measurements of microfluidic flow fields and suspended particles. In particular, two main needs are emerging in this domain: tools that can effectively be operated by non-expert users like biologists or physicians, and automated, real-time methods suitable for active force and flow control (e.g. to allow single-cell manipulation~\cite{gao2019recent}). Since the first application of microscopic PIV \cite{santiago1998particle}, about two decades ago, 
 several velocimetry and particle tracking methods have been proposed as diagnostic tools for microfluidics, both in 2D and 3D, using different principles such as defocusing~\cite{willert1992three,pereira2000defocusing}, astigmatic aberration~\cite{cierpka2010simple,cierpka2011calibration}, evanescent waves~\cite{zettner2003particle}, holography~\cite{hinsch2002holographic}, and tomography~\cite{kim2012comparison}. However, most of these methods require complex calibration procedures as well as experienced users to properly perform a measurement and are therefore not accessible to a wider community neither for usage nor for further development~\cite{cierpka2012particle}.

One method with the potential to meet these needs is the General Defocusing Particle Tracking (GDPT) which was proposed by Barnkob \etal~\cite{barnkob2015general} and is illustrated in Fig.~\ref{fig:intro}. The general requirements for the GDPT method is an optical system with small depth of field (particle images must have different shapes depending on their depth positions) and a set of reference images showing the particle image shapes at different known positions. Both requirements are typically fulfilled in microfluidic applications, where large magnification objective lenses are used and where the calibration images can easily be obtained by a systematic scanning of the microscope focus.  Moreover, distortions in microscopic objectives are typically small and the defocusing patterns are consistent across the entire camera sensor~\cite{barnkob2020general}. GDPT can indifferently be used on brightfield, darkfield, or fluorescent images as long as the image contrast is sufficiently high. Furthermore, in a standard GDPT algorithm, such as the free-to-use GDPTlab~\cite{GDPTlab}, most outliers (i.e. false positive) can be easily filtered out based on a single similarity parameter that evaluates how well a target image is matched to the calibration images. Therefore, and due to its simplicity, GDPT is receiving an increasing interest in microfluidics and lab-on-a-chip communities~\cite{taute2015high,liu2019investigation,boyko2020nonuniform}. Here in particular within the acoustic manipulation of microparticles~\cite{barnkob2018acoustically,karlsen2018acoustic,volk2018size,Qiu2019,van2020gradient,bode2020microparticle}, where information about the three-dimensional acoustophoretic behavior is crucial to further development as well as for the translation to industrial and clinical use, where feedback control is essential to secure stable and viable conditions~\cite{shaglwf2019acoustofluidic,nguyen2020acoustofluidic}.

In this work, we present a new algorithm to perform GDPT measurements in a fast, automated and versatile fashion. Its key feature is a segmentation-free approach to identify particles and estimate their 3D position. This detection step is followed by a refinement and iteration step to improve the accuracy and the identification of overlapping particles. The performance of the algorithm is evaluated using the benchmark dataset provided in Ref.~\citenum{barnkob2020general} with respect to error, valid detected particles, and processing time. Following, the algorithm is validated on experimental images with distortions and aberrations for the tracking of the three-dimensional motion of tracer particles inside an evaporating drop. Furthermore, we created a simulation framework to test the algorithm for use in real-time applications. The simulation is created using a combination of synthetic images~\cite{rossi2019synthetic} and analytical predictions of particle trajectories in acoustofluidic devices~\cite{muller2013ultrasound}.

\section{Algorithm description}
\seclab{fast}

\begin{figure*}[t!]
    \centering
    \includegraphics[width=\textwidth]{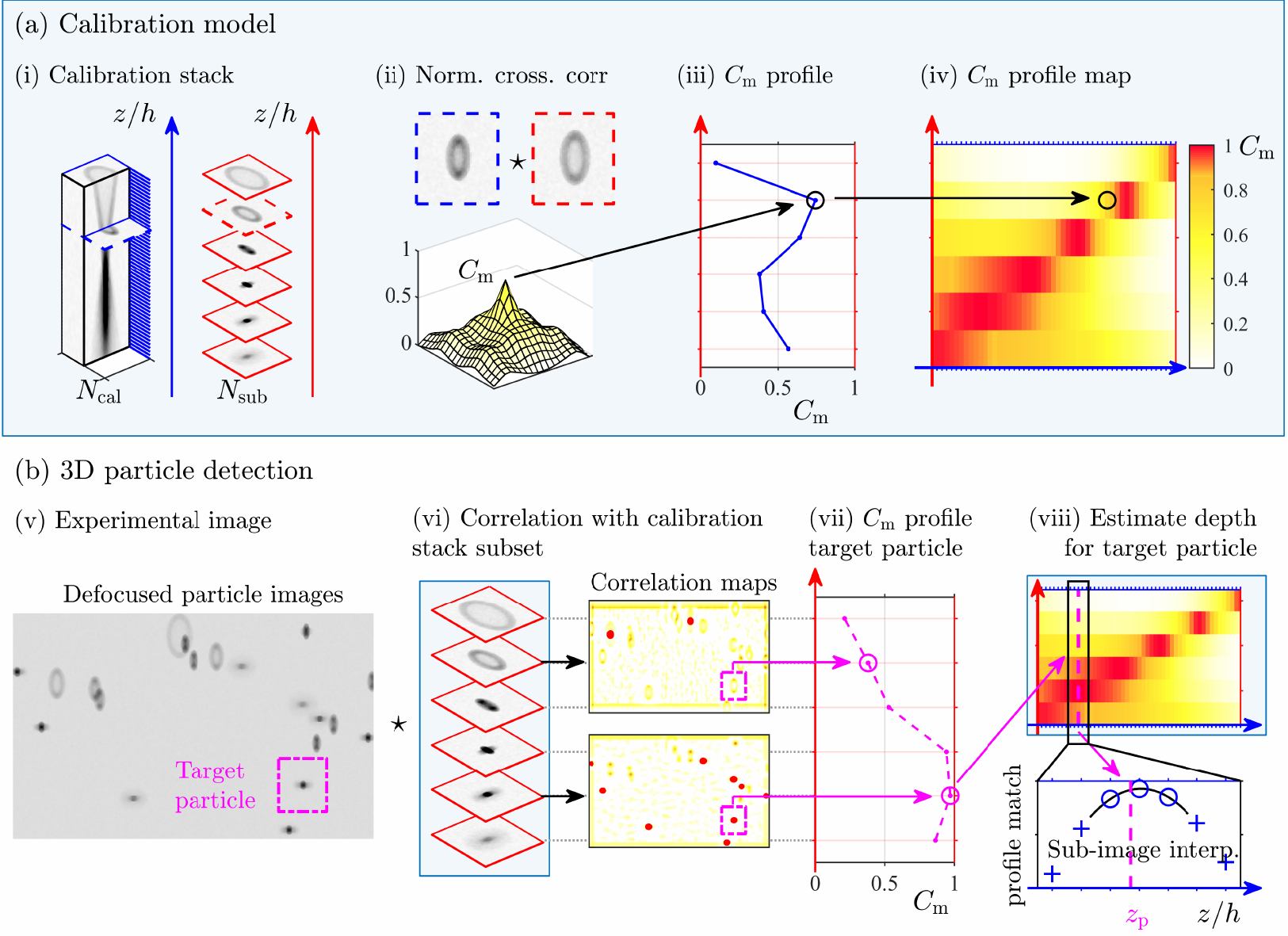}
    \caption{ Overview of the detection step. (a) The calibration model is based on a single calibration stack of $\Ncal$ single-particle calibration images. (i) A subset of $\Nsub$ calibration images is extracted. (ii) The similarity between two particle images is evaluated through the parameter $\Cm$, which represents the (local) peak value of the normalized cross-correlation ($\star$) between two images. (iii) By cross-correlating a given calibration image (dashed blue line) with the $\Nsub$ calibration subset images, a down-sampled $\Cm$ prediction profile across the entire measurement height is created. (iv) A $\Cm$ prediction map is created by creating down-sampled $\Cm$ profiles for each $\Ncal$ calibration images in the stack. (b) 3D particle detection scheme. (v) Input image of defocused particle images. (vi) The input image is cross-correlated with the $\Nsub$ calibration subset images resulting in $\Nsub$ correlation maps. Candidate particles are detected from the correlation maps from the local peaks larger (red areas) than a certain $\Cm$ threshold (shown for $\Cm=0.5$). (vii) For a subject candidate particle (dashed magenta lines), the $\Cm$ profile is extracted. (viii) The profile for the subject particle is matched to the $\Cm$ profile prediction map to determine its depth position with discrete precision $h/\Ncal$. Following, the continuous depth position $z_\mathrm{p}$ is determined through a three-point estimator to the three best profile map matches.}
	\label{fig:method}
\end{figure*}

The essential elements of the GDPT method is the { calibration model consisting of a set of calibration images} that maps defocused (or astigmatic) particle images with their respective depth position $z$ and a methodology to compare the similarity between a target particle image and the calibration images. Typically, the set of calibration images is obtained experimentally by taking subsequent images of a reference particle which is displaced at known positions, \eg\ using a motorized focusing stage while observing a sedimented particle, see \figref{intro}(b)~\cite{barnkob2015general,barnkob2020general}. In Refs.~\citenum{barnkob2015general} and \citenum{barnkob2020general}, and in this work, we use a calibration model based on calibration images of a single particle (calibration stack) and the normalized cross-correlation function \cite{lewis1995fast}. The normalized cross-correlation is used to rate the similarity between target and calibration images, using its maximum peak value as the similarity coefficient, referred to as $\Cm$, see \figref{method}(a,ii). The values of $\Cm$ can range from 0 to 1, with 1 corresponding to a perfect match between the target image and a calibration image.

A conventional approach for GDPT measurements consists of the following steps~\cite{barnkob2015general,GDPTlab}:
\begin{itemize}
    \item identify candidate particles using image segmentation,
    \item for each candidate particle, identify the best matching calibration image through $\Cm$ using an optimized iterative procedure,
    \item for each candidate particle, determine the $z$ position with "sub-image" resolution through interpolation in $(z,\Cm)$.
\end{itemize}
This procedure is very accurate, however, it is relatively slow since it needs to compute a large number of cross-correlations. For instance, the optimized approach in GDPTlab~\cite{GDPTlab} requires on average  average 6-10 cross-correlations for each candidate particle to converge to the best match. Moreover, the segmentation procedure must be optimized for each image type and fails if the background or the illumination is not uniform, therefore a pre-processing step is often required.

In this work, we present an improved algorithm to perform GDPT measurements based on the following steps\footnote{
The algorithm constitutes "Method 1" in \GDPTprogram, which is an open-source GDPT implementation published under the open-source license and available through \href{https://defocustracking.com/}{https://defocustracking.com/}.}:
\begin{description}
    \item[Detection step] fast and segmentation-free detection of particles with a robust first estimation of the 3D particle position,
    \item[Refinement step] for each candidate particle, refine the depth position walking through the calibration images stack,
    \item[Sub-image step] for each candidate particle, determine the $z$ position with "sub-image" resolution  using a three-point interpolation scheme,
    \item[Iteration step] iteratively repeat the preceding steps, while shading the images of detected particles, to increase the number of detected particles with overlapping images.
\end{description}

The detection and sub-image steps are the base of the algorithm and provide already a complete measurement. The subsequent steps improve the quality of the measurement but at a larger computational cost. The user can decide which strategy is most suitable for the application. All steps are described in details in the following.

\subsection{Detection step: Segmentation-free particle detection and fast particle position estimation}

The detection step is summarized in~\figref{method}. To prepare the calibration model, we first select a subset of $\Nsub$ calibration images from the total $\Ncal$ images in the calibration stack, see \figref{method}(a). For each of the $\Ncal$ calibration images, we calculate an "expected" $\Cm$ profile by performing a normalized cross-correlation with the $\Nsub$ images in the subset. This gives an \emph{a priori} expected $\Cm$ profile mapping, which are used during the evaluation for a fast identification of the particle $z$ positions. The detection step starts by performing normalized cross-correlations between the full image and the $\Nsub$ calibration images in the subset, resulting on a total of $\Nsub$ correlation maps, see \figref{method}(b). The correlation maps have values between 0 and 1, with their peak values located in the center of the particle images with shapes similar to the calibration image used to create the map. The center positions of candidate particles are identified collecting the locations of all peaks in the correlation maps with a value above a certain threshold (typically 0.5). Now, for each given candidate particle, the corresponding $\Cm$ profile across the $\Nsub$ number of $z$ positions is immediately available by reading the values of the $\Nsub$ correlation maps at that location (\figref{method}(b,vii)). Each correlation profile can following be compared with the mapping of expected $\Cm$ profiles to obtain a robust guess of the $z$ position. This is much faster than comparing 2D calibration images since the $\Cm$ profiles are 1D arrays of $\Nsub$ numbers. With this approach we have three significant advantages:
\begin{enumerate}
    \item Automatic detection: The in-plane and out-of-plane positions of candidate particles are determined by setting a single parameter, namely $\Cm$.
    \item Fast detection: Only $\Nsub$ cross-correlations are needed, regardless of the number of particles in the image.
    \item Versatility: The same procedure can be applied for any type of particle images (fluorescent, brightfield, with non-uniform illumination, etc.) without any additional steps.
\end{enumerate}

\begin{figure*}[t]
    \centering
    \includegraphics[width=\textwidth]{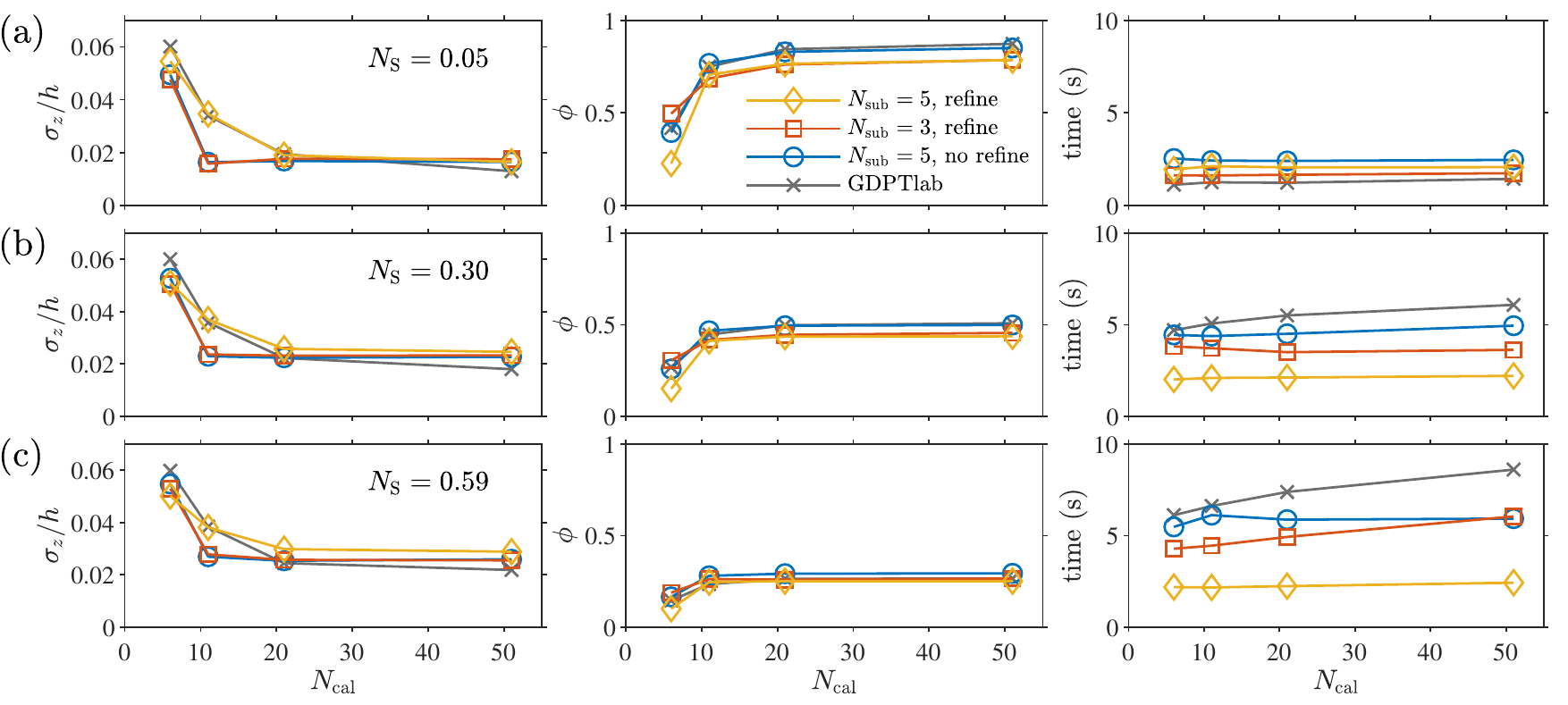}
    \caption{Parametric assessment of the presented algorithm { on synthetic images} in terms of error in the depth coordinate determination $\errz$, the relative number of valid detected particles $\detectedp$, and the processing time, as a function of the number of images in the calibration stack $\Ncal$, the number of images in the calibration subset $\Nsub$, with or without the refinement step. The assessment is performed on synthetic reference images for three values of the particle image concentration: (a) $\imconc = 0.05$, (b) $\imconc = 0.30$, and (c) $\imconc = 0.59$~\cite{barnkob2020general}. Results performed with the software GDPTlab are reported as reference.}
    \label{fig:errors-times}
\end{figure*}

\subsection{Refinement, sub-image, and iteration steps}

The preceding detection step provides a complete measurement with the identification of the particles and a robust estimation of their three-dimensional position. 

\textbf{Refinement step.} The detection step might not necessarily find the best matching calibration image and a refinement can be needed. All particles identified in the detection step is ordered according to their $\Cm$ values, from larger to smaller. Then, a walking procedure is applied to the first particle image to find the best match with the images in the calibration stack, searching in positions close to the $z$ value predicted by the first step. When the refined $z$ position is obtained, the particle image is removed from the image by replacing its area with the intensity values of the background. The same procedure is repeated for the second particle and so forth. At the end of the refinement step, all the detected particles have been removed from the image as illustrated in \figref{errors-iterations}(a).

\textbf{Sub-image step.} The determination of the $z$ position, both after the detection step and the refinement step, is produced with discrete output among the $\Ncal$ images separated by $h/\Ncal$. In order to obtain a continuous output with a ``sub-image'' resolution, the best matching calibration image must have been detected. Consequently, its neighbouring images are included for match interpolation, similar to what is typically done in digital PIV evaluations to obtain sub-pixel resolution, see \figref{method}(b). We use here a three-point parabolic estimator~\cite{Raffel2018}. 

\textbf{Iteration step.} As shown in~\figref{method}(b), in the detection step, the cross-correlation already filters out particles based on their image shape, i.e. only particle images similar to the calibration images are detected (highlighted in red). This allows to identify overlapping particles of different shapes. However, when the degree of overlapping between two particle images is too large, the height of the correlation peak decreases significantly and it is no longer possible to identify them. A way to improve the detection in this case is to blank out one of the overlapping particle images~\cite{wieneke2012iterative,liberzon2004xpiv}. This is done in the refinement step with the remaining particles images being those that were not identified in the first step. In order to increase the number of detected particles, the preceding steps are repeated, see \figref{errors-iterations}.

The above steps improve the accuracy and number of detected particles, however more computation time is consumed, therefore it might not be suitable for real-time applications. The relation between accuracy, number of detected particles, and computational time is investigated in the next section.

\begin{figure*}[t]
    \centering
    \includegraphics[width=\textwidth]{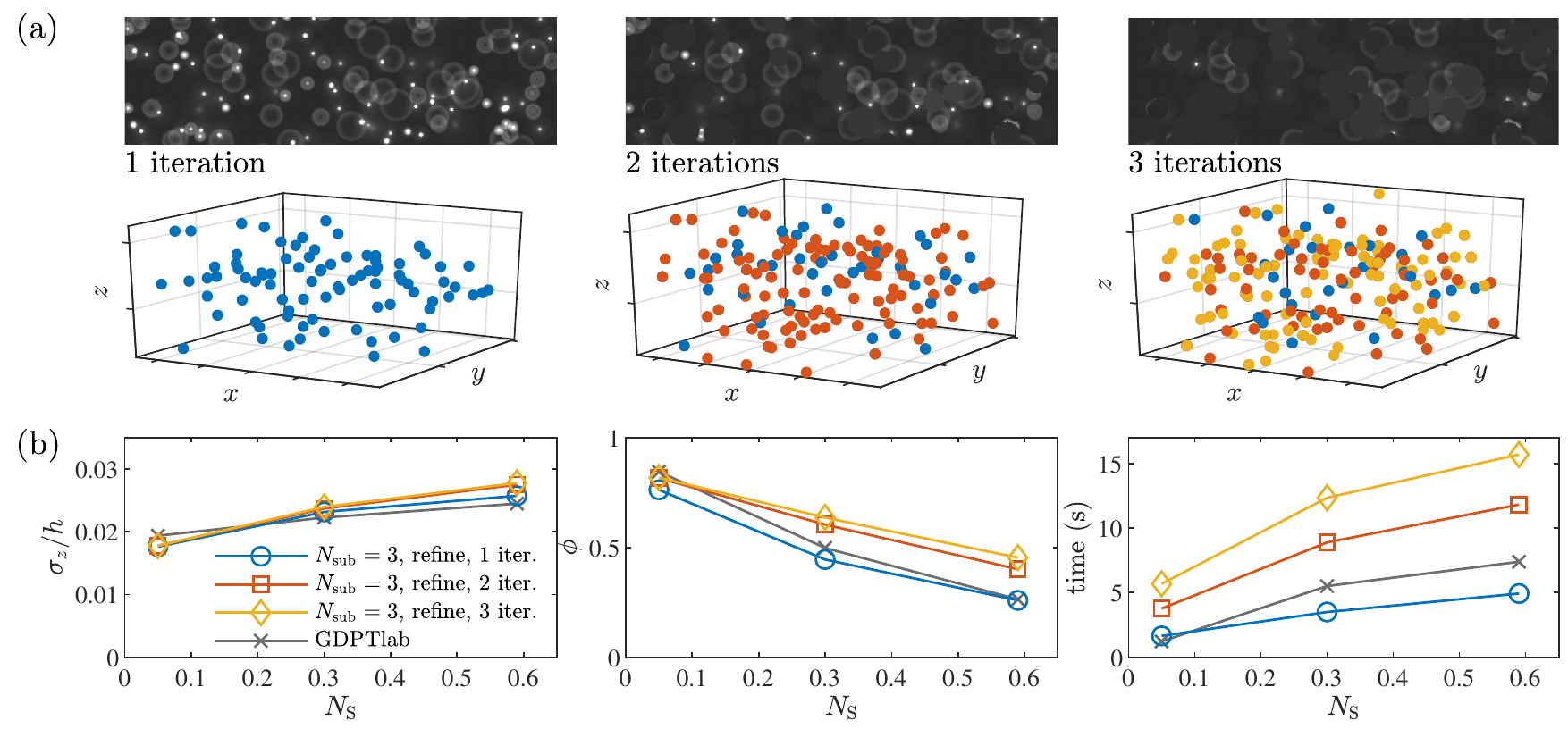}
    \caption{Performance assessment of the iteration step on synthetic images. (a) Portions of the analyzed images and reconstructed particle positions at each iteration. At the end of each iteration, the detected particle images are deleted, and the following iteration determines a new set of particle positions. (b) Error in the depth coordinate determination $\errz$, the relative number of valid detected particles $\detectedp$, and processing time as a function of the image concentration $\imconc$. The analysis was performed using $\Ncal = 21$, $\Nsub = 3$, and the refinement step, and for one, two, and three iterations. Results performed with the software GDPTlab are reported as reference.}
    \figlab{errors-iterations}
\end{figure*}

\section{Performance assessment on synthetic images}
\seclab{performance}

The presented algorithm allows to automatically detect particle images and determine their 3D position from a suitable calibration image set. Only few parameters must be chosen from the user before starting the evaluation, namely: 
\begin{itemize}
    \item The number of images $\Ncal$ in the calibration stack.
    \item The number of images $\Nsub$ in the calibration subset.
    \item The threshold value of $\Cm$ for identifying candidate particle images. 
    \item {Refinement step or no refinement step.}
    \item {The number of iterations.}
\end{itemize}
For the purpose of automated and real-time measurements, the most relevant parameters to evaluate are the uncertainty in the depth coordinate determination $\errz$, the relative number of valid detected particles $\detectedp$ (relative to the total number of particles in the image), and the overall processing time for each image. We do not consider here the uncertainty in the in-plane directions $\sigma_{x,y}$, which are normally one or more orders of magnitude less than $\errz$.  Measured particles with a $z$-error larger than 0.1 are considered outliers and not counted as valid detected particles. For a complete description of the assessment of GDPT methods, see Ref.~\citenum{barnkob2020general}. 

In order to assess the effect of different parameter settings on the algorithm, we used a standardized dataset of synthetic defocused particle images presented in Ref.~\citenum{barnkob2020general}. The images are created using MicroSIG, a synthetic image generator, based on ray tracing, for defocused and astigmatic particle images~\cite{rossi2019synthetic}. The dataset simulate measurements performed on 2-$\upmu$m-diameter particles, with 10$\times$ magnification over a measurement depth of 86 $\upmu$m. In particular, we used Dataset II, which contains sets of images of different particle image concentrations\footnote{The dataset can be downloaded through \href{https://defocustracking.com/}{https://defocustracking.com/}}. { The images are gray scale, 16-bit images with size of $1024\times1024$ pixels.} Following Ref.~\citenum{barnkob2020general}, we define {the particle image concentration in terms of  source density $\imconc$~\cite{adrian1984scattering,westerweel2000theoretical}, namely,} the number of particles in the image multiplied by the particle image area and divided by the total image area (or, alternatively, particle per pixels times average particle image area). 

Three parameter configurations have been investigated as a function of using different number of calibration images $\Ncal$: Configuration 1 with $\Nsub = 5$ and with refinement step (blue circles), Configuration 2 with $\Nsub = 3$ and with refinement step (red squares), and Configuration 3 with $\Nsub = 5$ and no refinement step (yellow diamonds). For all the configurations we used a single iteration step and a similarity threshold value $\Cm = 0.5$. The choice of $\Cm$ is mostly dictated by the user needs for a given experiment: Larger $\Cm$ values yield lower uncertainty at the cost of less detected particles and the other way around. The number of detected particles has also an impact on the computational time, as will be discussed later. Overall, different values of $\Cm$ do not play a significant role in the trends observed in the following investigation, therefore we show the results for one single $\Cm $ value. For reference, we performed the analysis also with a previous GDPT software, GDPTlab~\cite{GDPTlab}. GDPTlab uses a user-defined segmentation step (based on an intensity threshold) for particle identification and a refinement step as described in \secref{fast} B for the $z$ determination.

The results are presented in~\figref{errors-times} for three different particle image concentrations ($\imconc = 0.05, 0.30, 0.59$). In general, a use of $\Ncal$ larger than 10 is required, however values larger than 20 do not improve the performance significantly. Also, the value of $\Ncal$ minimally affect the processing time as the $z$ estimation in the detection step is done through simple matching of 1D profiles with $\Nsub$ points, see \figref{method}(b,viii). The refinement step  does not decrease the uncertainty significantly, but it results in a significantly larger processing time, which increases proportionally with the particle image concentration. On the other hand, without the refinement step (Configuration 3), the processing time is basically independent of the particle image concentration. All the three configurations provide approximately the same number of valid detected particles, which is however strongly affected by the particle image concentration. At $\imconc = 0.59$, the relative number of detected particles drops to around 30~\%. 

It is possible to increase the number of valid detected particles by adding more iterations of the refinement step as shown in~\figref{errors-iterations}. In particular, a second iteration brings this value up to 50~\%, whereas a third iteration does not improve it much further as the additionally-detected particles have similarity values below the threshold of $\Cm=0.5$. The cost of more iterations in terms of processing time is however significant.

Finally, in comparison with GDPTlab, the results are similar in terms of uncertainty and detected particles, showing that the segmentation-free approach is in this case as good as a conventional segmentation procedure based on an intensity threshold. However, the current analysis was performed on synthetic images with uniform background and no noise. The advantage of the segmentation-free approach is more evident on experimental images with noise and time-varying background, as shown in the next section. In terms of processing time, GDPTlab is strongly affected by the number of processed particles. Although it is faster at lower concentrations, it slows down rapidly as the number of particle increases, whereas the new algorithm allows a faster processing also for larger concentrations.

\begin{figure*}[t]
    \centering
    \includegraphics[width=1\textwidth]{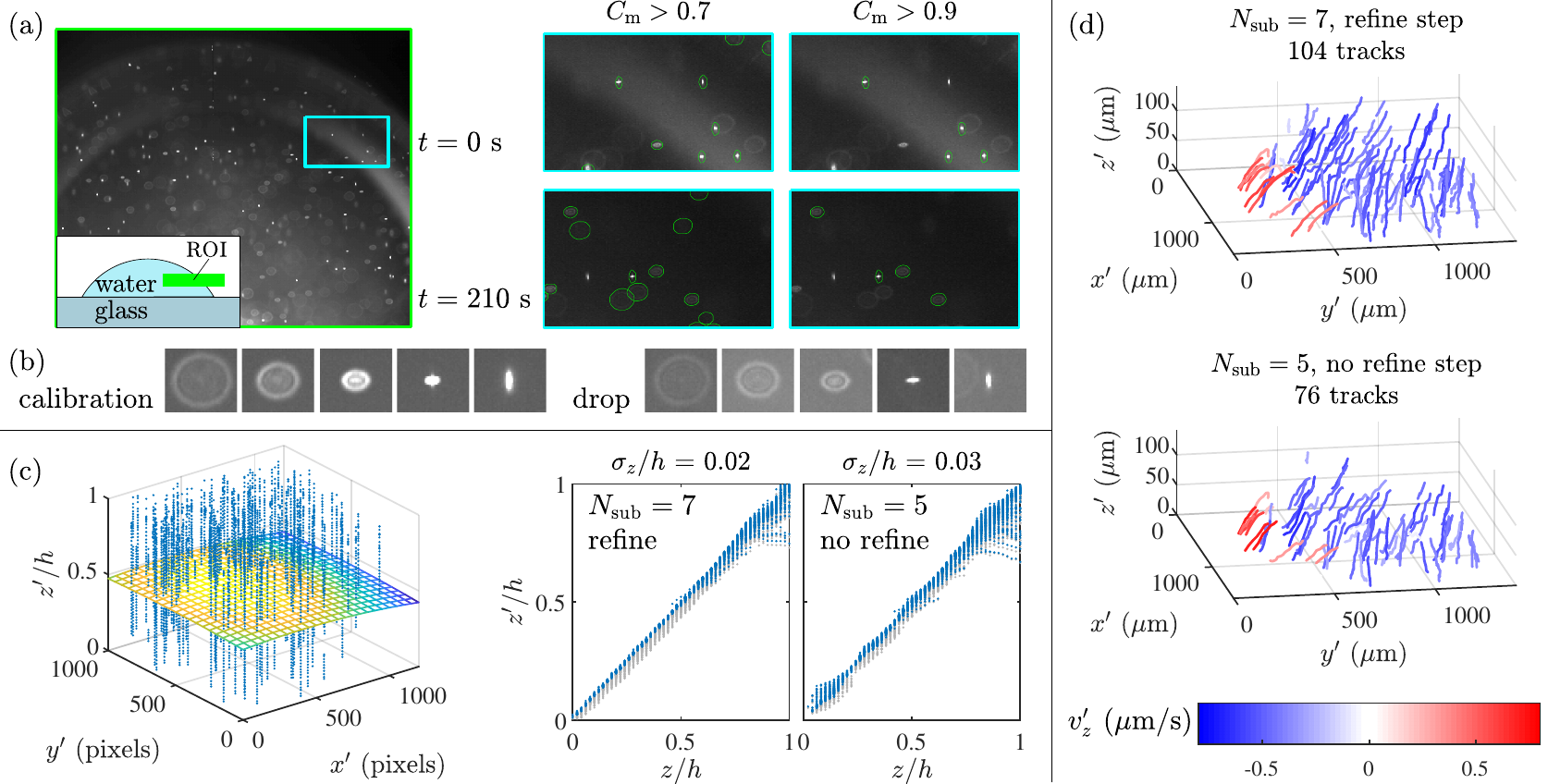}
    \caption{Application to experimental images: Particles inside an evaporating drop. (a) Example raw image showing the defocused images of fluorescent particles inside the region of interest (ROI, green area in sketch inset). The image subset inside the cyan-marked rectangle is shown for two frames at time $t=0$ s and $t=210$ s, respectively. For each frame, the detected particles are marked (green shapes) for $\Cm>0.7$ and  $\Cm>0.9$. No image pre-processing was used to treat the non-homogeneous background (due to reflections at the drop interface). (b) Particle image shapes in the calibration images and in the experimental images. (c) Uncertainty assessment and correction of systematic errors through evaluation of the calibration images obtained by taking images of sedimented particles at known depth positions $z$ (illustrated in Panel (b) in \figref{intro}). The measured ($x'$,$y'$,$z'$)-positions (blue dots) are used to map the systematic error due to the field curvature (face-colored surface). The uncertainty $\sigma_z/h$ of the measured depth coordinate $z'$ is accessed through $(z,z')$ shown for the normal (left) and the fast (right) GDPT processing as well as with (blue points) and without (grey points) the correction of the systematic error from the field curvature. (d) Measured ($x'$,$y'$,$z'$) particle trajectories inside the evaporating drop taken at 0.1 fps over a time of 210 s when using the normal (top) and the fast (bottom) GDPT processing.}
    \label{fig:drop}
\end{figure*}

{
\section{Application to experimental images}

To demonstrate the algorithm on experimental images, we chose images of tracer particles inside an evaporating droplet. We will not discuss the physics of the problem here or the details of the experimental setup, and for this we refer to Ref.~\citenum{rossi2019interfacial}. As a quick overview, a 2-mm-diameter droplet of pure water was deposited on a transparent glass slide. The drop was seeded with 1-$\upmu$m-diameter polystyrene spheres that were observed from the bottom using an inverted epi-fluorescent microscope. This example is interesting to test the proposed algorithm, since the image background is not uniform and it changes frame after frame due to reflections from the moving air-water interface (as shown in \figref{drop}(a)). Such images would need a pre-processing step for background removal in order to be processed with a conventional segmentation approach, such the one used in GDPTlab.   The current segmentation-free algorithm can be applied on raw images, without any pre-processing step.

The evaluation was performed on 16-bit images, with size of $1280\times1080$ pixels. The calibration stack was obtained on sedimented particles that were scanned moving the microscope objective focus with steps of 2 $\upmu$m across a depth of 100 $\upmu$m ($\Ncal = 51$). For the calibration, we chose one particle in the center of the image. We should briefly mention two practical aspects here. First, during the calibration, the objective lens moves in air and the particles are fixed, whereas during the measurement the objective lens is fixed and the particles moves in water. Although this has a minimal effect on the shape of the particle images, as shown in \figref{drop}(b), in order to obtain the same defocusing effect, the distance that a particle must travel in water is different from the distance that the objective lens has travelled in air. In this case, this difference is well-approximated by a pre-factor equal to the refractive index of water 1.33~\cite{rossi2012effect}. Second, in experimental images the defocusing patterns might change across the image sensor due to distortions, perspective errors, or other aberrations. These are systematic errors that can in most case be corrected, more information are given in Ref.~\citenum{barnkob2020general}. 

To correct for bias errors in the case at hand, we evaluated the positions of all the particles in the calibration images, which are particles lying on a flat surface and stuck to fixed in-plane positions.  As shown in \figref{drop}(c), particles across the entire sensor could be measured, perspective or distortion errors are negligible ($<0.2$ pixels), and the only relevant source of error is given by the curvature of the focal plane which is not perfectly flat (shown by the curved mesh in the figure). 

We can use these data also to assess the measurement uncertainty. In particular, we used two settings of the presented algorithm: $\Nsub = 7$, with refinement step (normal), and $\Nsub = 5$, without refinement step (fast). In \figref{drop}(c), we show the measured versus true $z$ positions for the two settings, the blue dots are the data after bias correction. Larger errors are observed in the top part of the calibration, where the images are dimmer and more blurred. Overall, the uncertainty difference between the two settings is small, and is about 2~\% for the normal setting and 3~\% for the fast setting (standard deviation of the error normalized on the total measurement depth~\cite{barnkob2020general}).

\begin{figure*}[t]
    \centering
    \includegraphics[width=.9\textwidth]{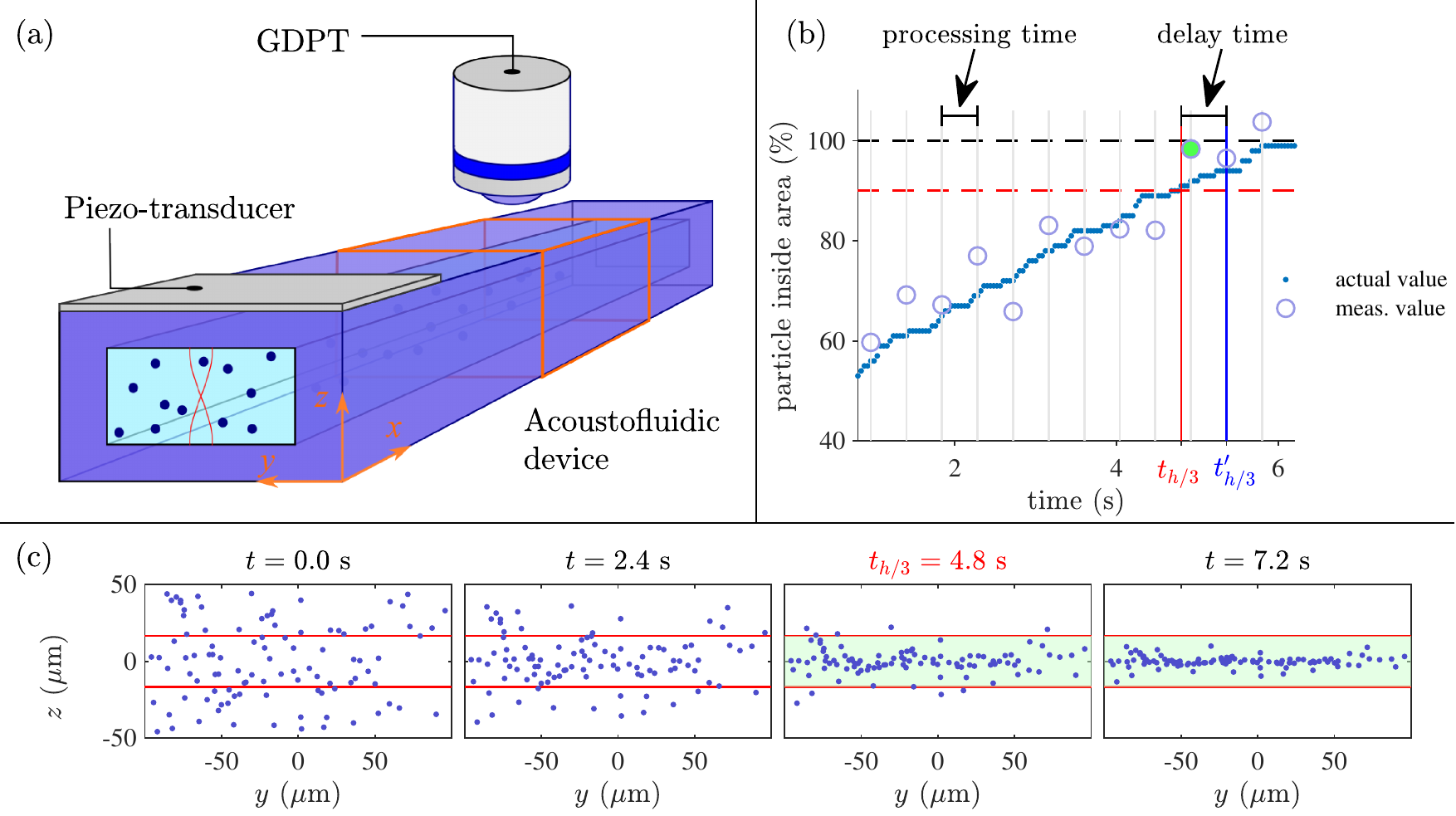}
    \caption{(a) Schematic of the acoustofluidic experiment object of the simulations: an acoustofluidic device is used to focus particles in the center height of a rectangular microchannel. Real-time GDPT measurements are used to monitor the position of the particles inside the channel and to identify the time $t_{h/3}$, when 90~\% of the particles are inside a vertical region of thickness $h/3$ (indicated by red horizontal lines in (c)). (b) Illustration of the percentage of particles that has reached the $h/3$ vertical region as a function of time. The points indicate the real number of particles inside the region, {the spacing in the time axes is determined by the frame rate of the acquisition. The circles indicate the number particle inside the region as measured by GDPT, the spacing in the time axis is determined by the computation time.} (c) Cross-sectional view of the simulated particle position in the measurement region for different time instants when assuming a Poiseuille flow with flow rate $Q = 3$ $\upmu$l/h and acoustic energy density $E_\mathrm{ac} = 0.3$ J/m$^3$.}
    \label{fig:sim}
\end{figure*}

Finally, we show the results of the droplet measurement. The images were acquired at 0.1 frames per second and processed with the two settings described above (normal and fast). The processing time per frame was 3 and 8 s, respectively, using a laptop computer with an Intel Core i7-7600U CPU processor running at 2.80 GHz with a RAM memory of 16 GB.  The particle trajectories and velocities where obtained using a simple nearest neighbor approach. No image pre-processing was used, and, as shown in \figref{drop}(a), the detection process worked properly independently of the background intensity fluctuations. The number of particles that are detected is determined by a threshold on their $\Cm$ value: A lower threshold catches more particles but increase also the measurement error and the presence of outliers.

The final particle trajectories using the two settings are reported in \figref{drop}(d), and correspond to a measurement volume of about $1600\times1300\times133$ $\upmu$m$^3$\footnote{Note that in Ref.~\citenum{rossi2019interfacial} a deeper and finer calibration stack was used, yielding a measurement volume of $1400\times1300\times150$ $\upmu$m$^3$ and depth uncertainty of 1 $\upmu$m.}. Particles on the bulk of the fluid move downwards toward the contact line, whereas particles at the interface move upward due to a thermally-induced Marangoni flow. Although the normal setting catches more tracks than the fast setting, the overall structures of the flow are well captured also by the fast setting. On the other hand, in case of real-time measurement, the fast setting allows a three times faster temporal resolution. 
}

\section{Simulation of a real-time microfluidic experiment}
\label{sec:acoustofluidic}

{In this section we set up a simulation framework to test the performance of the presented algorithm on a real-time application. We use simulations since we need to have access to the true particle positions to assess the algorithm performance. As a real-time application, we chose particle focusing in acoustofluidic devices, which is receiving a growing interest in biomedical applications~\cite{antfolk2019acoustofluidic}. }

{\textbf{Simulation of acoustofluidic focusing.}} We set up a simulation of a typical acoustofluidic experiment for separation or focusing of particles or cells~\cite{lenshof2012acoustofluidics,adams2012high}, see \figref{sim}(a). The simulation consists of 5-$\upmu$m-diameter polystyrene particles suspended in water in a microchannel of rectangular cross-section (width $w = 200$ $\upmu$m, height $h = 100$ $\upmu$m). The particle suspension is transported through the microchannel by an externally-driven flow and the resulting stream-wise particle velocities $u^\mathrm{flow}_x(y,z)$ are calculated from the analytical solution of a Poiseuille flow in a rectangular channel with flow rate $Q = 3~\upmu$l/h~\cite{bruus2008theoretical}. The microchannel is acoustically actuated by an attached piezo-electric transducer to induce a vertical half-wave standing acoustic waves of frequency $f_z = \cwa/(2h)=7.49~$MHz across the channel height. {The acoustic actuation leads to acoustic radiation forces on the suspended particles~\cite{karlsen2015forces} (we neglect any effects from acoustic streaming and particle-particle/particle-wall interactions~\cite{barnkob2012acoustic})} and through balance of the viscous drag force and the acoustic radiation force, we can calculate analytically the cross-sectional acoustophoretic particle velocities as
\begin{align}
    u^{\mathrm{rad}}_z(z) = \frac{2\pi}{3}\frac{a^2}{h}\frac{\Phi}{\eta} E_\mathrm{ac}\,\sin\bigg[2\pi\bigg(\frac{z}{h}+\frac{1}{2}\bigg)\bigg],
\end{align}
where $\eta$ is the fluid viscosity, $\Phi$ the acoustic contrast between particle and suspending fluid, and $E_\mathrm{ac}$ is the acoustic energy density. We used an acoustic energy density $E_\mathrm{ac} = 0.3$ J/m$^{3}$, which is a low but realistic value corresponding to maximum acoustic particle velocities of approximately 8~$\upmu$m/s~\cite{barnkob2010measuring,muller2013ultrasound}. Before the onset of the experiment, the particles are randomly distributed inside the channel and the resulting cross-sectional particle positions over a time of 8 seconds are shown in Fig.~\ref{fig:sim}(c).

{\textbf{Generation of synthetic images.}}
The acoustofluidic simulations were used to create synthetic images of the particles using MicroSIG~\cite{rossi2019synthetic}. The main setting for the synthetic images were a 10$\times$/0.3 objective lens plus astigmatic aberration on a 512$\times$512 pixels sensor (pixel size of 6.5 $\upmu$m{, 16 bit}). Gaussian noise was added to the images resulting in signal-to-noise ratio SNR for individual particle images ranging from 30 to 240. A typical synthetic image is shown in \figref{synth}(a), which corresponds to a classical darkfield image used in micro-PIV setups (i.e. fluorescent particles observed with an epi-fluorescent microscope). Additionally, we used two other types of images: Brightfield images obtained by inverting the values of the darkfield images (\figref{synth}(b)) and brightfield images with an added intensity disturbance, introduced to simulate non-uniform backgrounds or non-uniform illumination (\figref{synth}(c)). The disturbance consists of a 2D sinusoidal pattern.

{\textbf{Simulation of real-time experiment.}}
An \emph{ad hoc} Matlab routine was written to simulate real-time GDPT measurements with a camera frame rate of 25 fps in a section of the microchannel as illustrated in \figref{sim}(a). The objective of the simulation is to identify the "trigger" time $t_{h/3}$ for which 90~\% of the particles have been focused in a vertical center region of thickness $h/3$ (marked with red horizontal lines in \figref{sim}(c)). There are mainly three parameters to consider for the assessment of real-time GDPT measurements: (1) the computational time of a single evaluation, (2) the accuracy of the measurement, and (3) the number of detected particles (this strongly depends on the particle concentration, since it is more difficult to process overlapping particles, see \secref{performance}). Improving points (2) and (3) leads to longer computational times therefore a trade-off must be found. The computational time sets also the temporal resolution of the real-time measurement, therefore a random delay time proportional to the temporal resolution is expected. This is shown in \figref{sim}(b), where the percentage of particles that has reached the $h/3$ vertical region is plotted as a function of time. The points indicate the real number of particles inside the region and the red vertical line indicates the real "trigger time" $t_{h/3}$. The circles show the corresponding GDPT measurements, which have lower temporal resolution due to the computational time of each frame. The blue vertical line marks the corresponding GDPT "trigger time" $t'_{h/3}$. In the hypothetical case of perfect GDPT measurements, the delay time $t'_{h/3}-t_{h/3}$ is between one and two times the temporal resolution. However, due to the GDPT measurement uncertainty, a smaller or larger delay time can occur. As an example, if the delay time is negative, GDPT detects that 90~\% of the particles have reached the region before they actual have.

\begin{SCfigure*}
\caption{The presented algorithm can be applied to {different} types of images without any additional processing steps, which are here shown by synthetic images generated with MicroSIG. The images are {synthetic images} of 5-$\upmu$m particles taken with a 10$\times$/0.3 objective lens and the use of astigmatic aberration. (a) Fluorescent image. (b) Bright-field image. (c) Bright-field image disturbed with a sinusoidal intensity pattern simulating a non-homogeneous background or illumination.}
\includegraphics[width=1.5\columnwidth]{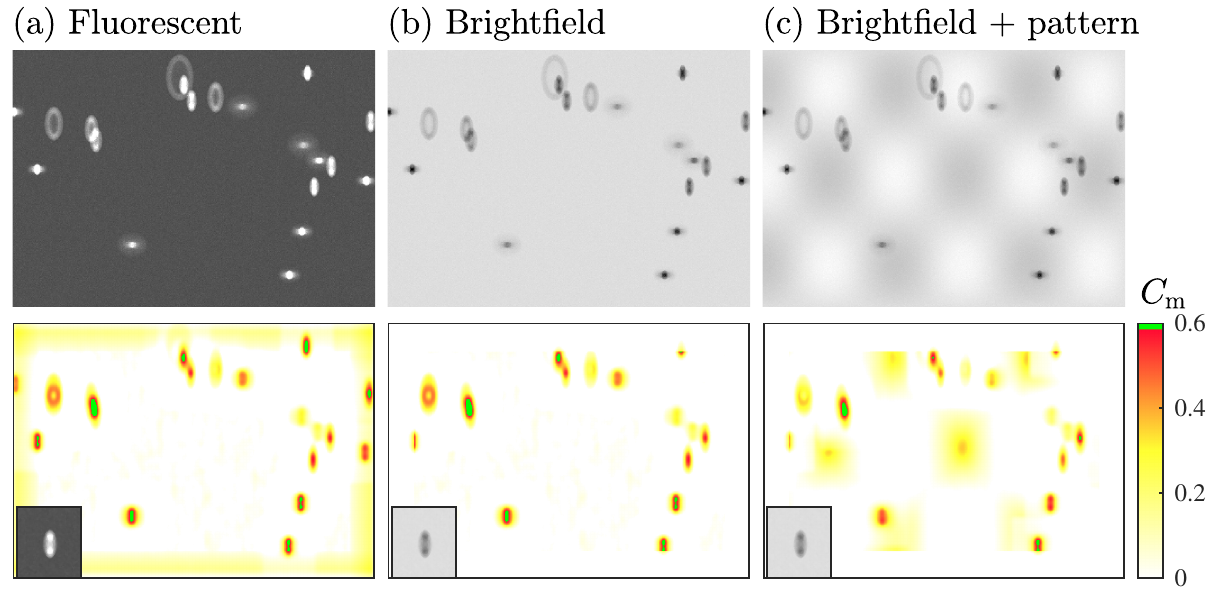}
\figlab{synth}
\end{SCfigure*}

{\textbf{Results of real-time GDPT measurement.}}
In a first set of experiments, we tested the performance of real-time GDPT measurements on the fluorescent images. As discussed in \secref{performance}, we test different strategies to decrease the computational time, namely by reducing the total number of images in the calibration stack $\Ncal$, by reducing the number $\Nsub$ of images in the subset of the calibration stack, and by removing the refinement step. In particular, see \figref{res}(a), we calculated the detection delay time $t_{h/3}-t'_{h/3}$ as a function of the average computational time for each image. The grey area indicates the expected results in the case of having no uncertainty in the GDPT measurements. The simulations show that removing the refinement step speeds up significantly the computational time with relatively low impact on the accuracy. On the other hand, decreasing the number of images in the calibration stack minimally decrease the computational time but can cause a failure of the measurement. The corresponding average percentage of the number of detected particles is shown in \figref{res}(b).

\begin{figure*}[t]
    \centering
    \includegraphics[width=.9\textwidth]{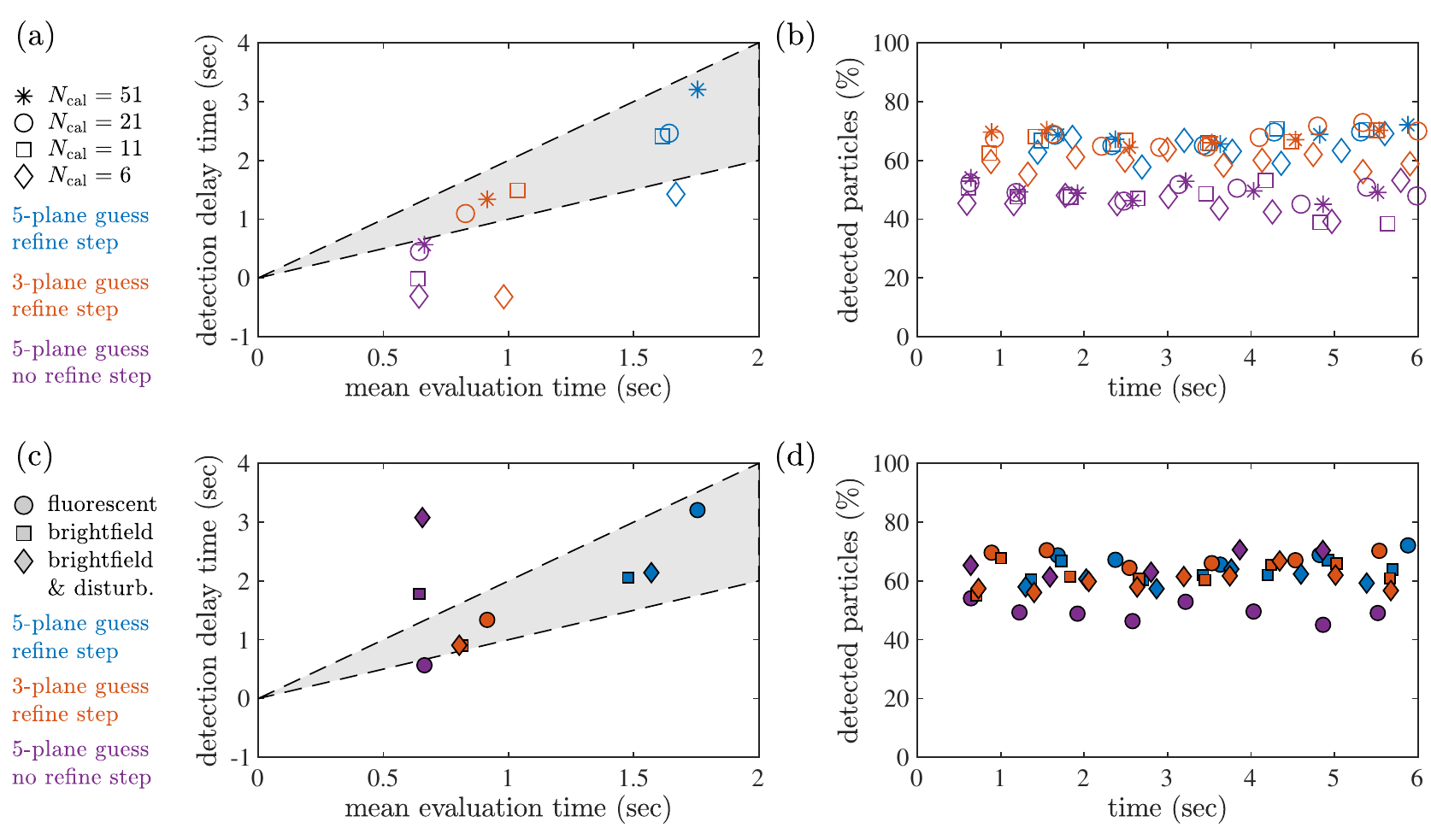}
    \caption{{ Simulation of a }real-time application of GDPT on a microparticle acoustophoresis experiment as shown in \figref{sim}. The results are shown for different algorithm settings and for (a-b) using different number of images in the calibration stack $\Ncal$ and (c-d) for different types of images with $\Ncal=51$. (a,c) Measured detection delay time $t_{h/3}-t'_{h/3}$ as a function of the mean evaluation time where the grey area represents the random delay expected for the given evaluation time. (b,d) Average percentage of detected number of particles as a function of time.}
    \figlab{res}
\end{figure*}

In a second set of experiments, we tested the performance of the presented algorithm when applied to different types of images: fluorescent, brightfield, and brightfield with disturbance, see \figref{synth}. This time we always used $\Ncal = 51$ images in the calibration stack. The results of the simulations are presented in \figref{res}(c-d) and show that no significant difference is observed when the refinement step is used, proving that the presented approach is suitable for different image types but that in most challenging situation the refinement step is needed.  

The simulations were performed on $512\times 512$-sized, {16-bit} images on a laptop computer with an Intel Core i7-7600U CPU processor running at 2.80 GHz with a RAM memory of 16 GB. With this setup it was possible to achieve an evaluation time of about one second.

\section{Conclusions}

We have presented a new algorithm for performing 3D particle tracking using GDPT that does not need preliminary pre-processing or segmentation steps. The presented algorithm needs to compute only few cross-correlations in comparison with other iterative approaches and is therefore suitable for a fast evaluation time. In addition, the algorithm allows for the detection of overlapping particles. The performance of the algorithm, in terms of uncertainty in the depth determination, detection of valid particles, and processing time, was tested on synthetic {and experimental images}.  The algorithm was tested for real-time application by setting up a framework for real-time simulation of an acoustophoretic experiment, created using synthetic images and an \emph{ad hoc} Matlab routine. The real-time simulations show, that even without using the refinement step, the presented algorithm is able to perform automated control-tasks and that work robustly on different types of images (darkfield or brightfield) also with fluctuations of the background intensity. The presented algorithm and simulation framework set the base for use of GDPT in real-time applications as well as for the further development and improvement hereof.

\begin{acknowledgments}
The research leading to these results has received funding from the European Union's Horizon 2020 research and innovation programme under the Marie Sklodowska-Curie grant agreement no. 713683 (COFUNDfellowsDTU).
\end{acknowledgments}


\end{document}